\documentstyle[prd,aps,eqsecnum,epsf,twocolumn]{revtex}
\begin{document}
 
\title{The Peculiar--Velocity--Field in Structure Formation Theories with 
Cosmic Strings}

\author{Carsten van de Bruck$^1$}
\address{\small\it
        $^1$Institut f\"ur Astrophysik und Extraterrestrische Forschung\\
        Auf dem H\"ugel 71\\
        53121 Bonn, Germany\\
        cvdb@astro.uni-bonn.de}

\date{today}

\maketitle

\begin{abstract}
We investigate the peculiar velocity field due to long cosmic strings in 
several cosmological models and analyse the influence of a nonscaling 
behaviour of the string network, which is expected in open cosmological models 
or models with a cosmological constant. It is shown that the deviation of 
the propability distribution of the peculiar velocity field from the normal
distribution is only weak in all models. It is further argued that one 
can not necessarily obtain the parameter $\beta=\Omega_{0}^{0.6}/b$ from 
density and velocity fields, where $\Omega_0$ is the density parameter 
and $b$ the linear biasing parameter, if cosmic strings are responsible 
for structure formation in the universe. An explanation for this finding 
is given. 
\end{abstract}


\section{Introduction}
To understand the origin and the formation of structure 
in the universe is one of the most challenging problems 
in modern cosmology. There are two competitive theories 
which try to explain the origin of the seeds. The first 
one is inflation, in which the universe undergoes an epoch 
of fast (inflationary) expansion, triggered by a scalar 
field, called the inflaton. Quantum fluctuations in the 
inflaton field are stretched on superhorizon scales which 
are turned into matter fluctuations at later times. 
These matter fluctuations represent the seeds of the observed 
structure today \cite{bi:Steinhardt}. In the other theory 
topological defects are responsible for structure formation. 
For example, cosmic strings might be produced in a phase 
transition in the very early universe. If they are 
heavy enough, they influence the cosmological fluid 
and could seed the structure in the universe (see 
\cite{bi:VileShell,bi:HindmarshKibble} for reviews and 
references).

There are several ways to test these theories. For example, 
they make different predictions for the angular fluctuation 
spectrum of the CMBR on small scales. The predicted spectra 
can then be compared with the data, for example of future 
projects such as MAP and PLANCK. Related to this is the 
comparison of the predicted matter power spectrum with the 
data. Cosmic strings could further be tested with 
``astrophysical'' tests, such as the expected gravitational 
radiation background from strings etc. It's interesting to 
note, that all these tests gave consistent results for the 
string parameter $\mu$, the mass per length on the string 
\cite{bi:VileShell,bi:HindmarshKibble}. 

Another possibility for testing structure formation theories 
was proposed by \cite{bi:Kofman}. The propability distribution 
of the peculiar velocity field should be different in 
inflationary models and models with topological defects 
such as cosmic strings. However, as emphasized by 
\cite{bi:Vachaspati,bi:Moessner}, the propability distribution 
of the peculiar velocity field in cosmic string theories is 
Gaussian to high accuracy. This conclusion was based on the 
assumption that the string network reaches a scaling behaviour.

It was shown by several authors that the scaling solution is 
only expected in the Einstein--de Sitter model 
\cite{bi:Martins,bi:vdBruck}; in open models, in flat models with a 
cosmological constant and in closed (loitering) models the 
behaviour of the network is different from scaling. Further, 
the transition to the matter scaling behaviour is much longer 
than previously estimated \cite{bi:MartinsShellard}. These 
possible sources of deviation from the string scaling solution 
should have interesting consequences. A first step was done 
in \cite{bi:albrecht} and \cite{bi:avelino}.
Whereas in \cite{bi:albrecht} it was shown, that only a drastic 
departure from scaling could solve the problems of structure 
formation which cosmic strings, \cite{bi:avelino} have shown 
that there is only weak dependence of the density parameter 
$\Omega_{0}$ in open and flat models with cosmological 
constant on the normalisation of $G \mu$ from COBE data. 
It might be, that the cosmic string scenario is successfull 
also in the absence of the scaling behaviour, i.e. that it 
works well in open models or in models with a cosmological 
constant. 

In this paper we investigate the influence of cosmic strings 
on the peculiar velocity field. Earlier investigations of the 
peculiar velocity field concentrated on the spectrum of the 
field, i.e. it´s dependence on the length scale $L$, see e.g. 
\cite{bi:Vachaspati,bi:vachiperi,bi:hara}. We are interested 
in the effects of a departure from the scaling behaviour. We 
use an approximation, first introduced by \cite{bi:Vachaspati} 
to calculate the effects of many strings. 

The paper is organized as follows: In section II we discuss the 
influence of cosmic strings on the peculiar velocity field. 
Our calculations of the string network are based on the 
calculations by \cite{bi:vdBruck}. Our results for the 
peculiar velocities are presented in section III. In section IV 
we argue that if cosmic strings seed the structure in the universe 
then the peculiar velocity field and the density field is correlated 
but one cannot obtain information on the parameter 
$\beta=\Omega_{0}^{0.6}/b$, where $\Omega_0$ is the density parameter 
and $b$ is a linear bias parameter. In section V we summarize 
our results and give some conclusions. Throughout the paper we 
set $c=1$. 

\section{The peculiar velocity field due to long cosmic strings}
The space--time of a straight cosmic string is similar to the 
Minkowski space--time, except for a deficit angle $\Delta \phi$, 
given by \cite{bi:VileShell,bi:HindmarshKibble}:
\begin{equation}
\Delta \phi = 8 \pi G \mu \gamma_{s} v_{s}.
\end{equation}
As a result, the matter gets a kick towards the plane swept out 
by the string ($v_{s}$ is the string velocity and $\gamma_{s}$ 
is the Lorentz factor). The velocity kick due to a wiggly 
string is given by
\begin{equation}\label{kick}
u_{s} = 4 \pi G \mu \gamma_{s} v_{s} f = 3.8 (G\mu)_{6} (\gamma_{s}
v_{s})f \mbox{ }{\rm km/s},
\end{equation}
with 
\begin{equation}
f =  1 + \frac{1}{2v_{s}^{2}\gamma_{s}^{2}}\left(1 
       - \frac{T}{\mu} \right).
\end{equation}
The term $f$ corresponds to the small scale structure on the 
string, where $\mu$ is the effective mass per unit length on 
the string and $T$ is the string tension. 

\subsection{The Zeldovich approximation}
To calculate the peculiar velocity field, we use the Zeldovich 
approximation, in which the physical coordinates of a 
particle are written by 
\begin{equation}\label{displacement}
{\bf r} ({\bf x},t) = a(t) \left[ {\bf x} + {\bf \psi}({\bf x},t)\right],
\end{equation}
where $a(t)$ is the scale factor, ${\bf x}$ is the comoving coordinate 
and ${\bf \psi}$ is the displacement vector due to inhomogenities in 
the cosmic fluid, i.e. cosmic strings in our context \cite{bi:zeldovich}. 
The equation of motion is given by 
Newtons law
\begin{equation}
\ddot{{\bf r}} = -\nabla_{{\bf r}} \Phi.
\end{equation}
The gravitational field is connected with the matter distribution 
(Poisson equation), which can be obtained from linearising Einsteins 
field equation:
\begin{equation}
\nabla_{{\bf r}}^{2} \Phi = 4\pi G(\rho_b + \delta \rho) + \Lambda c^2.
\end{equation}
In this equation $\rho_b$ is the matter density, $\delta\rho$ the matter 
density fluctuation and $\Lambda$ the cosmological 
constant. To first order one obtains
\begin{equation}
\delta \equiv \frac{\rho - \rho_{b}}{\rho_{b}} = 
- \nabla_{{\bf x}}\cdot\psi({\bf x},t), 
\end{equation}
Where $\rho$ is the total matter density. This leads to 
\begin{equation}
\nabla_{{\bf r}}^{2} \Phi = 4\pi G \rho_{b}(1-\nabla_{{\bf x}}\cdot\psi) 
+ \Lambda c^2
\end{equation}
Integration of this equation and substitution of 
\begin{equation}
\ddot{{\bf r}} = \frac{\ddot a}{a} {\bf r} + 2\dot{a}\dot{{\bf \psi}} 
+ a \ddot{{\bf \psi}}
\end{equation}
and the second Friedmann equation
\begin{equation}
\frac{\ddot a}{a} = -\frac{4\pi G}{3}\rho_{b} + \frac{\Lambda c^2}{3} 
\end{equation}
leads to the evolution equation for the displacement ${\bf \psi}$ 
\cite{bi:zeldovich}:
\begin{equation}\label{zeldovich}
\ddot{\vec\psi} + 2\frac{\dot a}{a}\dot{\vec\psi} - 4\pi G \rho_{b}\vec\psi 
= 0.
\end{equation}
For our purposes we have to calculate the peculiar velocity field, which 
can be obtained from eq. (\ref{displacement}):
\begin{equation}\label{pefield}
{\bf v}_{pec} = a\dot{\bf \psi}.
\end{equation}
The effect of the cosmological constant on the evolution of a 
density perturbation is only due to the effect of $\Lambda$ on the 
evolution of the scale factor $a$. 

\subsection{The influence of cosmic strings}
We use an analytical approximation (the so called {\it multiple 
impulse approximation}), first introduced by Vachaspati 
\cite{bi:Vachaspati}, which also was successfully applied to calculate 
the CMBR anisotropies \cite{bi:peri}. We 
devide the time interval from $t_{eq}$ 
(at which structures starts to form) to $t_{0}$ in $N$ steps with 
$t_{i+1}=2t_{i}$. Between $t_{i}$ and $t_{i+1}$ the strings 
intercommute, form loops etc. so that (approximately) at $t_{i+1}$ 
the ``new ordered state'' of the network is uncorrelated with the 
``old state'' at $t_{i}$. Again, at this time the network influences 
the matter within the horizon (due to scalar field radiation 
scales larger than the horizon are not affected). This is not 
true in vacuum dominated 
epochs. In this case the velocity of the strings decreases and 
therefore the propability of string interaction decreases. This 
means that the new state {\bf is} correlated with the old one. 
However, in such epochs the number of strings within the Hubble 
horizon decreases rapidely and therefore our results aren't 
changed significantly (see below). 

At $t_1 = t_{eq}$ each string within the horizon gives the 
matter a kick in the direction of the surface swapped out by 
the string:
\begin{equation}
{\bf v}_{1,i} = u_{\rm s}{\bf k}_{i,1}, 
\end{equation}
where ${\bf k}_{i,1}$ is a (random) unit vector in direction of the string $i$.
The resulting peculiar velocity from all strings at $t_1$ is 
\begin{equation}
{\bf v}_{1} = \sum_{i=1}^{n_{g,1}} u_{\rm s}{\bf k}_{i,1}, 
\end{equation}
The sum is now taken over the number $n_{g,1}$ of all strings within the 
horizon at $t_1$. This peculiar velocity field grows between $t_1$ and $t_2$ 
by a factor ${\cal A}(t_1,t_2)$ via eq. (\ref{zeldovich}) and 
eq. (\ref{pefield}):
\begin{equation}
{\cal A}(t_i,t_f) = \frac{a(t_f)|\dot{\bf \psi}(t_f)|}{a(t_i)|\dot{\bf \psi}(t_i)|}.
\end{equation}
At the time $t_{2}$ the peculiar velocity field is given by 
\begin{eqnarray}
{\bf v}_{2} &=& {\cal A}(t_1,t_2){\bf v}_{1} 
+ u_{s}\sum_{j=1}^{n_{g,2}}{\bf k}_{2,j} \nonumber \\ 
&=& u_{s} \left[\sum_{i=1}^{2}\sum_{j=1}^{n_{g,i}} 
{\cal A}(t_{2},t_{i}){\bf k}_{i,j}\right]. 
\end{eqnarray}
Here we have used that ${\cal A}(t_i,t_j){\cal A}(t_j,t_l) 
= {\cal A}(t_i,t_l)$ and that the velocity of the strings is the same in every 
epoch. For our purposes this is a good approximation, because when the 
strings slow down the number of strings within the horizon also decreases.

Iteration leads to the peculiar velocity field at the present 
time on a scale $L$:
\begin{equation}\label{basic}
{\bf v}_{0}(L) = u_{\rm s}\left[\sum_{i=1}^{N_L} 
\sum_{j=1}^{n_{g,i}} {\cal A}(t_{0},t_{i}){\bf k}_{i,j}\right].
\end{equation}
In this equation $N_L$ is the number of Hubble time steps during which a 
volume of comoving size $L^3$ experiences string impulses, $n_{g,i}$ is the 
number of strings within the horizon at the time $t_i$. We assume that the 
vectors ${\bf k}_{i,j}$ are random, that is
\begin{equation}
<{\bf k}_{i,j} \cdot {\bf k}_{l,m}> = \delta_{il}\delta_{jm}.
\end{equation}
From these equations we calculate the RMS velocity numerically on a scale 
$L(t_{eq})$. On scales smaller than $L(t_{eq})$ the peculiar velocity field 
depends only weakly on $L$ whereas on scales larger than $L(t_{eq})$ the 
predicted velocity field scales as $L^{-1}$ \cite{bi:vachiperi}. 

\subsection{Network parameter}
We use the calculation from \cite{bi:vdBruck} for the 
statistical properties of the string network. We set the number of 
strings within the horizon $H^{-1}$ by 
\begin{equation}\label{ansatzns}
n_s = 1 + (\xi\cdot H)^{-1},
\end{equation}
where $\xi$ is the characteristic length scale of the string network, 
defined by
\begin{equation}
\rho_{\infty} = \frac{\mu}{\xi^{2}},
\end{equation}
where $\rho_{\infty}$ is the density of the long strings. In the radiation 
dominated epoch $n_s$ is about 10, in the (late) matter dominated epoch 
(with scaling) this number is about 3. In more general cosmological models 
this number is a function of time \cite{bi:vdBruck}. Later we will 
discuss the influence of the ansatz (\ref{ansatzns}).

\section{Results}
We calculate the peculiar velocity field for four representative models, 
shown in Tab. 1. For the Einstein--de Sitter (E--dS) model we discuss the 
influence of the long transition between the radiation and matter scaling 
solution \cite{bi:MartinsShellard}. As a test, we include the case for an 
ideal scaling in the E--dS model. The propability distributions of the 
peculiar velocity fields at a scale $L(t_{eq})$ for the models are shown in 
Fig. 1--5. Each plot was obtained after 50000 realisations. 
For an exact scaling behaviour in the E--dS model this distribution was shown 
to be Gaussian \cite{bi:Vachaspati,bi:Moessner}. We obtain the same result 
(see Fig.1). 
In the case of the long transition between the radiation and matter scaling 
behaviour the distribution remains nearly Gaussian. However, 
the distribution becomes broader and the peculiar velocity increases 
(Fig. 2). The Gaussian character of the propability distribution can 
be found in the other models, too. There is only a slight deviation at 
large and small velocities. 

In the models, we obtain a peculiar velocity at a 
scale corresponding to $L(t_{eq})$ given by (the length scales 
are calculated with $H_0 = 100$km/(s$\cdot$Mpc)):
\begin{eqnarray}\label{result1}
v_{pec}(L_{eq}\approx 70\mbox{ }{\rm Mpc})_{\rm closed} \\ \nonumber
 = (460 \pm 200)(G\mu)_{6} (\gamma_{s}v_{s})f\mbox{ } {\rm km/s},
\end{eqnarray}
\begin{eqnarray}
v_{pec}(L_{eq}\approx 1\mbox{ }{\rm Mpc})_{\rm EdS,ni} \\ \nonumber 
 = (1740 \pm 760)(G\mu)_{6} (\gamma_{s}v_{s})f\mbox{ }{\rm km/s},
\end{eqnarray}
\begin{eqnarray}
v_{pec}(L_{eq}\approx 1\mbox{ }{\rm Mpc})_{\rm EdS,id} \\ \nonumber
= (1240 \pm 570)(G\mu)_{6} (\gamma_{s}v_{s})f\mbox{ }{\rm km/s},
\end{eqnarray}
\begin{eqnarray}
v_{pec}(L_{eq}\approx 10\mbox{ }{\rm Mpc})_{\lambda,{\rm flat}} \\ \nonumber
= (280 \pm 120)(G\mu)_{6} (\gamma_{s}v_{s})f\mbox{ }{\rm km/s},
\end{eqnarray}
\begin{eqnarray}\label{result2}
v_{pec}(L_{eq}\approx 10\mbox{ }{\rm Mpc})_{\rm open} \\ \nonumber
= (80 \pm 35)(G\mu)_{6} (\gamma_{s}v_{s})f\mbox{ } {\rm km/s}.
\end{eqnarray}

The length scale corresponding to the time $t_{eq}$ is set to be  
$0.1 H(t_{eq})^{-1}$ \cite{bi:Vachaspati}:
\begin{equation}
L_{eq} \approx 1.1 \frac{1}{\Omega_{0}} h_{0}^{-2} {\rm Mpc}
\end{equation}
Note, that in \cite{bi:Vachaspati} the length scale was set to 
be $0.7t_{eq}$. Therefore, we assume a somewhat pessimistic view 
when strings could significantly influence the volume at $t_{eq}$. 
In our picture, the volume must be within the typical length scale 
between all strings. However, the volume is influenced by the strings 
outside the volume and therefore we somewhat underestimate the 
peculiar velocities. However, this can be taken into account with including 
a parameter $\zeta$, which modifies our ansatz (\ref{ansatzns}) 
(see below, eq. (\ref{zeta})).
 
Within this length the peculiar velocity remains nearly constant, because 
a smaller length corresponds to times $t<t_{eq}$ in which perturbation 
grow only weakly. This would imply that for the closed model we would expect 
nearly constant bulk flows on scales smaller than 70 Mpc, which is indeed 
observed. The situation in the other models is not so clear, because for 
scales larger that $L_{eq}$ the velocity decreases as $L$ increases 
($v \propto L^{-1}$).

It is interesting to note that in {\it all} models the standard deviation 
is related to the mean value by
\begin{equation}
\sigma \approx 0.45 v_{\rm mean}.
\end{equation}

For our calculations we have used the ansatz (\ref{ansatzns}) for the number 
of strings within the horizon. Although this should be a good approximation 
we could set  $n_s =\zeta(1+(\xi\cdot H)^{-1})$. The frequency distribution 
remains Gaussian, however, now the RMS velocity and the standard deviation 
is given by 

\begin{equation}\label{zeta}
v_{pec,\zeta} = \sqrt{\zeta}v_{pec,\zeta = 1}
\end{equation}
and
\begin{equation}
\sigma_{\zeta} = \sqrt{\zeta} \sigma_{\zeta = 1}.
\end{equation}
Here, $v_{pec, \zeta = 1}$ and $\sigma_{\zeta = 1}$ is given by 
eq. (\ref{result1}--\ref{result2}) for the cosmological models. 
The peculiar velocities therefore depend on the parameter 
\begin{equation}\label{alpha}
\alpha = \sqrt{\zeta}\mu_{6}(v_{s}\gamma_{s}) f.
\end{equation}

To conclude, {\it the fluctuation of number of strings doesn't change 
the shape of the propability distribution of the peculiar velocities 
and the amplitude depends on the same set of parameters (\ref{alpha}) 
as in the case for a ideal scaling behaviour of the string network.} 
However, the effective number of strings and the maximum length on 
which coherent bulk flows are expected, depends on the cosmological 
parameters $\Omega_{m,0}$, $\lambda_{0}$ and $H_0$.

\section{Matter distribution, bulk flows and biasing}
The results presented in the last section imply that the 
parameter $\beta = \Omega_{m,0}^{0.6}/b$ could not be obtained 
from velocity--density reconstruction methods such as POTENT. To 
see this, we remember that the fundamental equation, on which 
these kinds of reconstruction methods are based, is given by 
\cite{bi:Peebles,bi:Dekel}
\begin{equation} \label{potent}
\nabla \cdot {\bf v} = -\beta H \delta.
\end{equation}
Here, $\delta$ is the density fluctuation. On the other side, 
the continuity equation holds:
\begin{equation} \label{conti}
\nabla \cdot {\bf v} = -\dot\delta.
\end{equation}
In fact, in linear approximation eq. (\ref{potent}) can be obtained 
from eq. (\ref{conti}). The important point is that if the ratio 
$\dot\delta/ \delta$ is independent of space, an arbitrary 
application of eq. (\ref{potent}) can lead to an under-- or 
overestimation of $\beta$ if one applies eq. (\ref{potent}) 
arbitrarily to the data. This was shown by \cite{bi:Babul} in the 
context of the explosion scenario. To demonstrate this point 
we repeat their short analytical example: 

Let us consider an empty universe with 
$\Omega_0 = 0$, filled with massless particles. At some time 
$t_i$, the matter gets a kick due to a cosmic string (in the paper by 
Babul et al., the case of the explosion scenario was 
considered, but in the case of cosmic strings the analysis is 
identical). The linear Euler equation reads:
\begin{equation}
\frac{\partial {\bf v}}{\partial t} + 2H(t){\bf v} = {\bf v}_{\rm string}
\delta (t-t_{i}).
\end{equation}
The density contrast evolves according to
\begin{equation}\label{dichte}
\frac{\partial^2 \delta}{\partial t^2} 
+ 2H(t)\frac{\partial \delta}{\partial t} = 0
\end{equation}
This equation can be solved with the boundary contitions at the time 
$t_i$, which are $\delta({\bf x},t_i)=0$ and $\dot \delta({\bf x},t_i) 
= \zeta({\bf x},t_i) = - \nabla_{\bf x}\cdot {\bf v}_{\rm string}
({\bf x},t_i)$. The solution of the equation (\ref{dichte}) is given by 
($a(t_i) = 1$)
\begin{eqnarray}
\delta({\bf x},t) & = &\zeta({\bf x},t_i)t_i(a(t) - 1)/a(t) , \\
t \dot\delta({\bf x},t) &=&\zeta({\bf x},t_i)t_i/a(t) .
\end{eqnarray}
One can see, that the ratio $\dot\delta/\delta$ is {\it independent 
of space}. We can use the continuity equation (\ref{conti}) and 
equation (\ref{potent}) to get
\begin{equation}
\beta_{\rm eff} = \frac{1}{a(t) - 1}.
\end{equation}
Although the true value is zero, an observer, applying (\ref{potent}) to 
the data, will get a value that is different from the true one. Only at 
late times $\beta_{\rm eff}$ will approach the value $\beta = 0$. 

We expect similar results for the models in table 1. The cosmological 
model changes the time dependence of $\delta$ and $\dot \delta$, but 
in general, if there were only one velocity kick on the matter, 
the ratio $\dot \delta / \delta$ would be independent of space. 

We have seen, that in cosmic string models of structure formation 
the observed peculiar velocity field is a vector sum of many contributions 
of strings. We write the general solutions of $\dot \delta$ and 
$\delta $, which arise from one kick as 
\begin{eqnarray}
\delta_{i,j}({\bf x}, t_i) &=& \zeta({\bf x}, t_i)_{i,j} {\cal B}(t,t_i),
\label{1}\\
\dot \delta_{i,j}({\bf x}, t_i) &=& \zeta({\bf x}, t_i)_{i,j} {\cal C}(t,t_i)
\label{2}.
\end{eqnarray}
The fields $\dot \delta$ and $\delta$, which arise from {\it all} 
strings in the past are given by 
\begin{eqnarray}
\dot \delta ({\bf x}, t)= \sum_{i=1}^{N}\sum_{j=1}^{n_{g,i}} \dot \delta_{i,j}({\bf x},t), \\
\delta({\bf x}, t) = \sum_{i=1}^{N}\sum_{j=1}^{n_{g,i}} \delta_{i,j}({\bf x}, t).
\end{eqnarray}
If we now replace $\zeta_{i,j}$ by $\delta_{i,j} / {\cal B}$, we can write 
\begin{eqnarray}
\dot \delta= \sum_{i=1}^{N}\sum_{j=1}^{n_{g,i}} \delta_{i,j}({\bf x}, t_i) 
\frac{{\cal C}(t,t_i)}{{\cal B}(t,t_i)}.
\end{eqnarray}
If we compare this with eq. (\ref{1}) we see that there is a relation 
between the velocity field and the density field, but not of the same 
form as in eq. (\ref{potent}). For cosmic strings we could write 
\begin{equation}\label{goal}
\nabla \cdot \vec v = - \sum_{i=1}^{N}\sum_{j=1}^{n_{g,i}} \delta_{i,j}
({\bf x}, t_i) \frac{{\cal C}(t,t_i)}{{\cal B}(t,t_i)}.
\end{equation}
If we compare eq. (\ref{goal}) with (\ref{potent}), we found that 
\begin{equation}\label{beta1}
\beta_{\rm eff} = \frac{1}{H_0}\frac{\sum_{i,j} \delta_{i,j} 
                      {\cal C}_{i,j}/{\cal B}_{i,j}}
                       {\sum_{i,j} \delta_{i,j}},
\end{equation}
where $\cal B$ and $\cal C$ are the solutions of the perturbation equations. 
This equation can be interpreted as follows: each kick gives an effective 
$\beta_{{\rm eff},i,j} = {\cal C}_{i,j}/{\cal B}_{i,j}$. Therefore, eq. 
(\ref{beta1}) is the weighted mean of the $\beta_{{\rm eff},i,j}$:
\begin{equation}
\beta_{\rm eff}=\frac{1}{H_0} \frac{\sum_{i,j}\delta_{i,j}
\beta_{{\rm eff},i,j}}{\sum_{i,j}\delta_{i,j}}.
\end{equation}
This represents a measure of the departure of an exact relation between the 
velocity and density field. 

\section{Discussion \& Conclusions}
In this paper we have considered the properties of the peculiar velocity 
field of galaxies in structure formation theories with cosmic strings. 
We considered the fact that the string network might not have 
developed a scaling behaviour (as is the case in open models or models 
with a cosmological constant) and showed that the propability distribution 
of the peculiar velocities is nearly Gaussian. The RMS peculiar velocity 
depends on the (effective) number of strings within the horizon and on the 
string parameter $G\mu$. The length, within the peculiar velocity is nearly 
independent of the scale, depends on the cosmological parameter 
$\Omega_0$, $\lambda_0$ and $H_0$. 

Open models have more problems with the amplitude of the peculiar velocity 
field. Only an unplausible high value of $f$ could solve the problems 
($f\approx 5$). The situation might be better in flat models with a 
cosmological contant. However, on scales larger than $L_{\rm eq}$ 
the peculiar velocity drops with $L^{-1}$, that is on a scale of about 60 Mpc 
we expect in model 4 peculiar velocities of 50--100 km/s. This is not in 
agreement with the observed value of 350--450 km/s. The situation is 
very good in the closed model. On scales smaller than $L_{eq}$ the 
velocity increases only weakly as $L$ decreases 
\cite{bi:Vachaspati,bi:vachiperi,bi:hara} and remains nearly constant at 460 km/s 
(for $ G\mu = 10^{-6}$ and $f \gamma_s v_s \approx 1$) up to 
scales of about 200 Mpc.

The results imply that if cosmic strings seed the structure in the 
universe, one can not necessarily obtain the density parameter from the 
data. Comparison of density fields and velocity fields lead to an 
effective value, which is a measure of the deviation of an exact relation 
between the velocity and density field. 

Further work should be done on structure formation with cosmic strings in 
order to investigate the effects on a non--scaling behaviour of the 
cosmic string network.

\vspace{0.5cm}
{\bf Acknowledgements:} I thank Matthias Soika and Harald Giersche 
for discussions. This work was supported by the Deutsche 
Forschungsgemeinschaft (DFG).

\mbox{ }

\begin{center}
\begin{table}
\begin{tabular}{|l|l|l|l|l|l|}
Model & $K$ & $\Omega_{0}$ & $\lambda_{0}$ & H$_{0}$ & $N_{eq}$ \\
\hline
1     & +1& 0.014 & 1.08 & 90 & 13 \\
\hline
2     & 0 & 1.0   & 0.0  & 60 & 20 \\
\hline
3     & --1& 0.1   & 0.0  & 60 & 14 \\
\hline
4     & 0 & 0.1   & 0.9  & 60 & 15 \\
\end{tabular}
\caption{The four representative cosmological models. $K$ is the curvature 
parameter, $\Omega_0$ is the matter density parameter, $\lambda_{0}$ is the 
cosmological term, $H_{0}$ is the Hubble parameter (in km s$^{-1}$Mpc$^{-1}$). 
$N_{eq}$ is the number of Hubble steps between $t_{eq}$ and $t_{0}$.}
\end{table}
\end{center}

\newpage
\mbox{ }
\newpage

\thispagestyle{plain}

\begin{figure}[htb] 
  \begin{center}
    \leavevmode
    \epsfxsize=13cm
    \epsffile{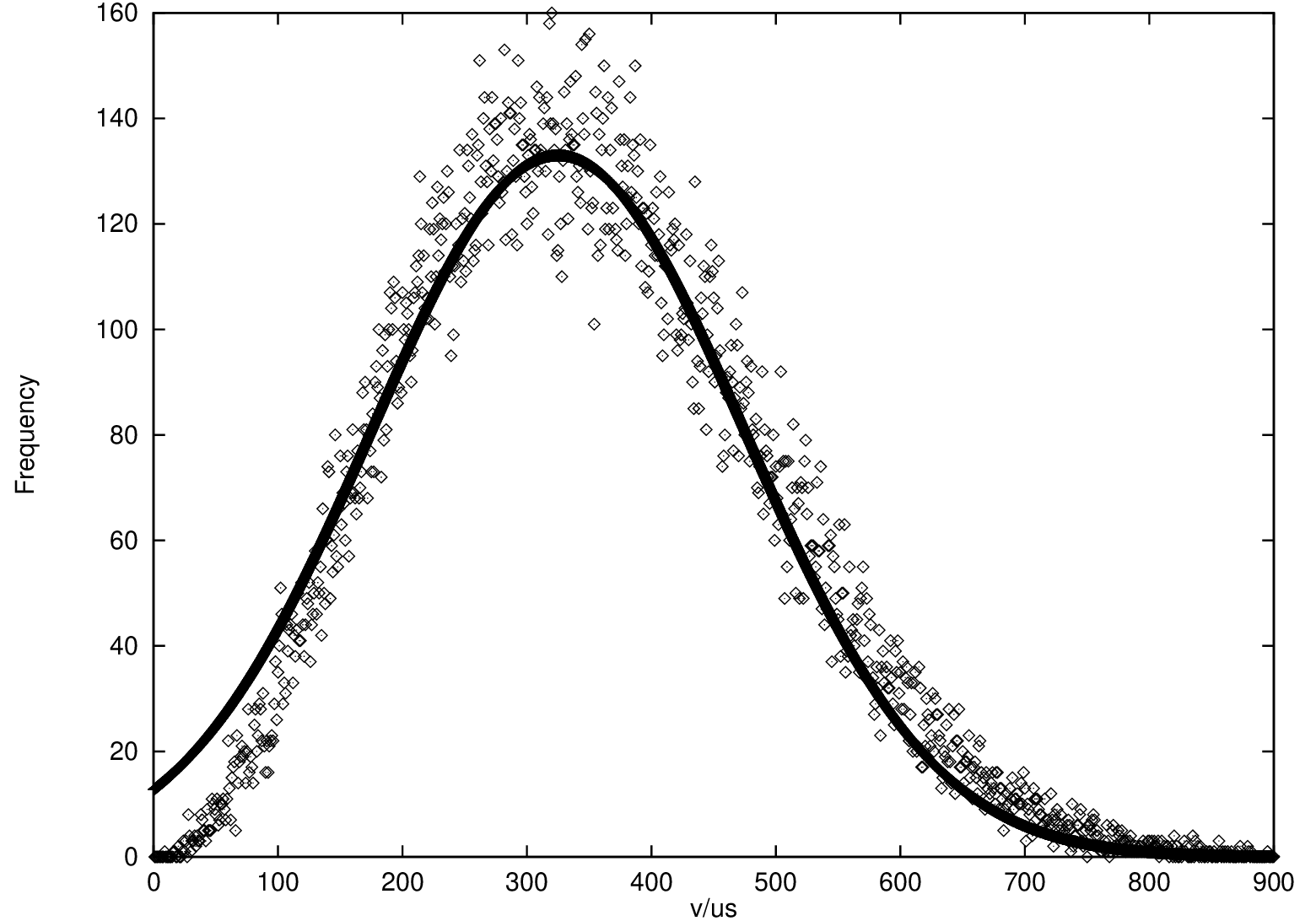}\vspace{0.5cm}
    \caption
    {The propability distribution of the peculiar velocity field at a scale 
     corresponding to $t_{eq}$ in the E-dS model with ideal scaling. The solid 
     curve is the normal distribution. The mean value of $v_{pec}/u_s$ 
     is given by 325 and the standard deviation by 150.}
    \label{fi1}
  \end{center}
\end{figure}

\newpage\mbox{ }\newpage
\thispagestyle{plain}

\begin{figure}[htb] 
  \begin{center}
    \leavevmode
    \epsfxsize=13cm
    \epsffile{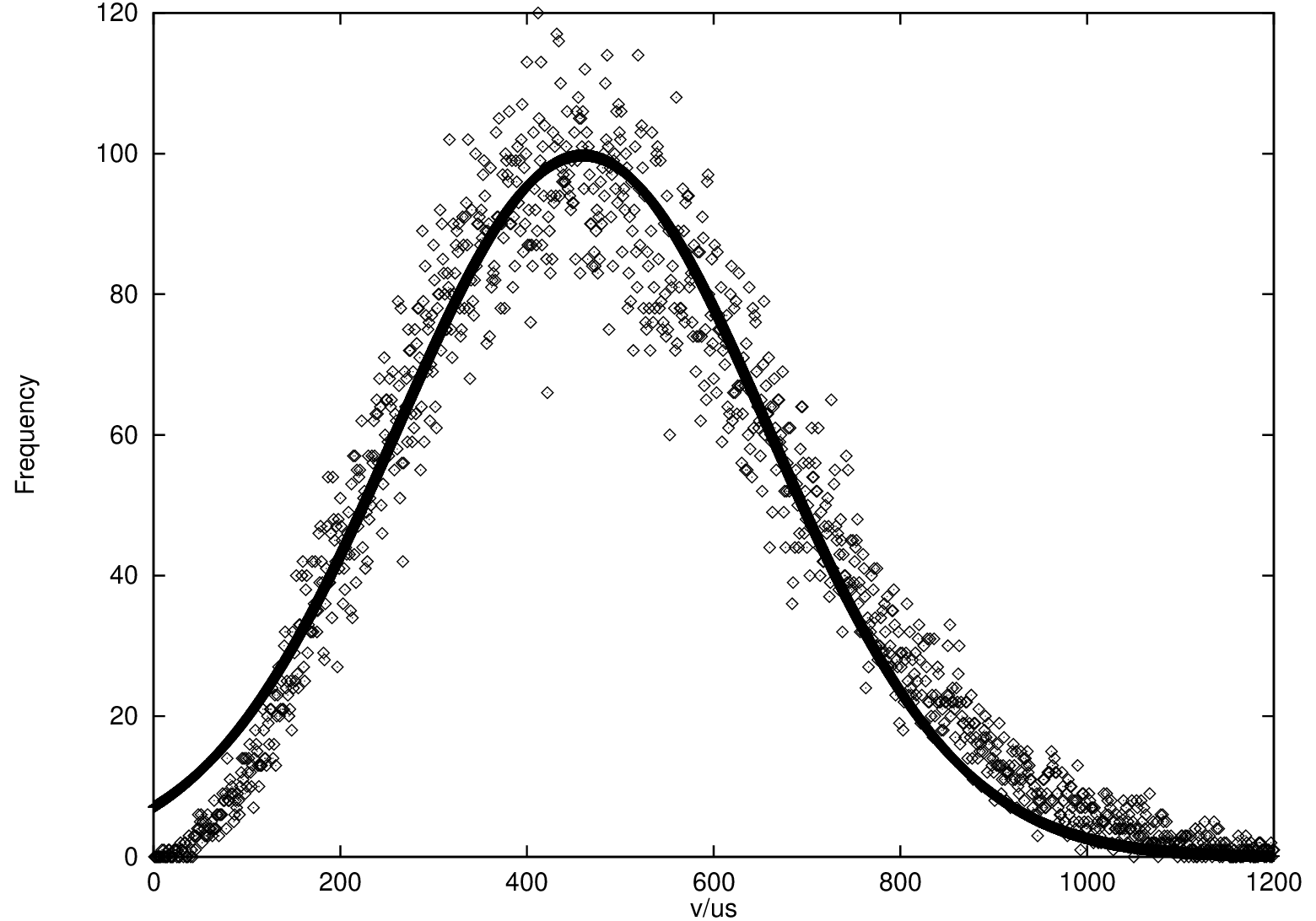}\vspace{0.5cm}
    \caption
    {The propability distribution of the peculiar velocity field at a scale 
     corresponding to $t_{eq}$ in the E-dS model with non-ideal scaling. The 
     solid curve is the normal distribution. The mean value of $v_{pec}/u_s$ 
     is given by 460 and the standard deviation by 200.}
    \label{fi2}
  \end{center}
\end{figure}

\newpage\mbox{ }\newpage
\thispagestyle{plain}

\begin{figure}[htb] 
  \begin{center}
    \leavevmode
    \epsfxsize=13cm
    \epsffile{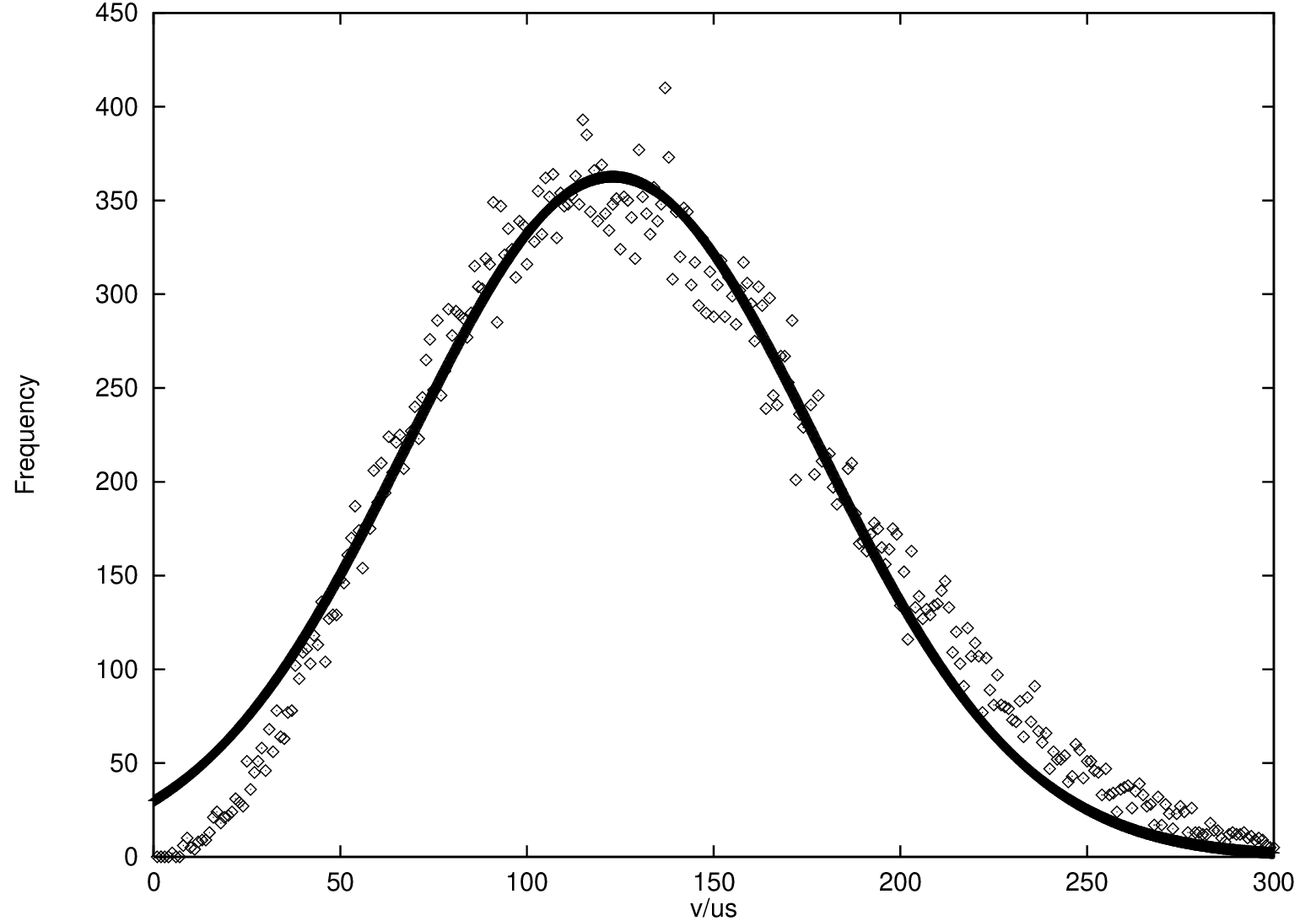}\vspace{0.5cm}
    \caption
    {The propability distribution of the peculiar velocity field at a scale 
     corresponding to $t_{eq}$ in the closed model. The solid curve is the 
     normal distribution. The mean value of $v_{pec}/u_s$ is given by 
     123 and the standard deviation by 55.}
    \label{fi3}
  \end{center}
\end{figure}

\newpage\mbox{ }\newpage
\thispagestyle{plain}

\begin{figure}[htb] 
  \begin{center}
    \leavevmode
    \epsfxsize=13cm
    \epsffile{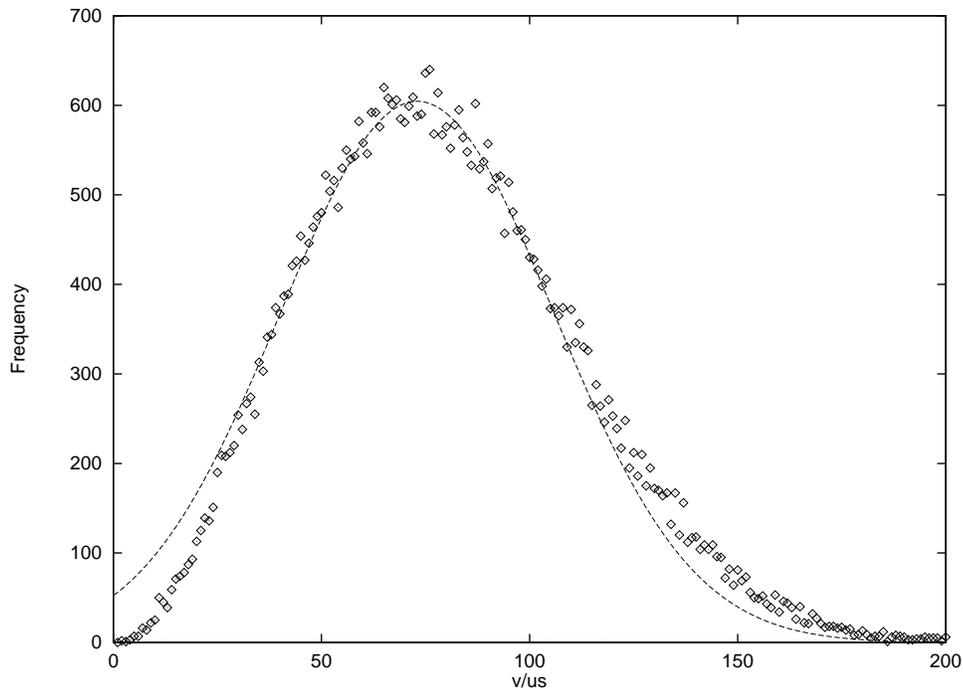}\vspace{0.5cm}
    \caption
    {The propability distribution of the peculiar velocity field at a scale 
     corresponding to $t_{eq}$ in the flat model with cosmological constant. 
     The solid curve is the normal distribution. The mean value of 
     $v_{pec}/u_s$ is given by 73 and the standard deviation by 33.}
    \label{fi5}
  \end{center} 
\end{figure}

\newpage\mbox{ }\newpage
\thispagestyle{plain}

\begin{figure}[htb] 
  \begin{center}
    \leavevmode
    \epsfxsize=13cm
    \epsffile{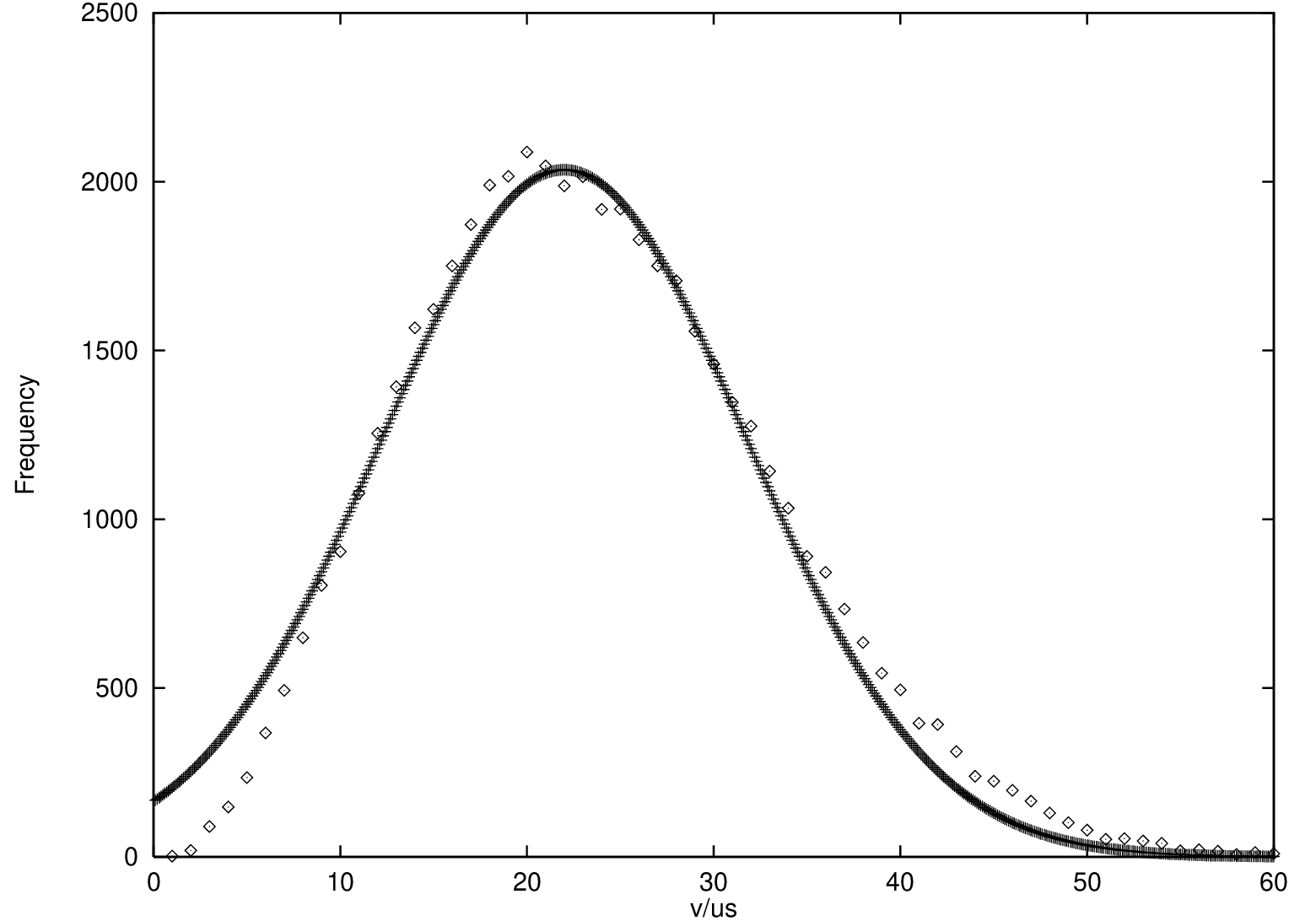}\vspace{0.5cm}
    \caption
    {The propability distribution of the peculiar velocity field at a scale 
     corresponding to $t_{eq}$ in the open model. The solid curve is the 
     normal distribution. The mean value of $v_{pec}/u_s$ is given by 
     22 and the standard deviation by 9.8. }
    \label{fi4}
  \end{center}
\end{figure}


\begin{thebibliography}{99}

\setlength{\baselineskip}{0.5cm}

\bibitem{bi:Steinhardt}
P.J. Steinhardt, in: {\it Particle and nuclear astrophysics and 
cosmology in the next millennium}, eds. E.W. Kolb \& R.D. Peccei, 
World Scientific (1995).

\bibitem{bi:VileShell}
A. Vilenkin and E.P.S. Shellard, {\it Cosmic strings and other topological 
defects} (Cambridge University Press, England, 1994).

\bibitem{bi:HindmarshKibble}
M. Hindmarsh and T.W.B. Kibble, Rep. Prog. Phys. {\bf 58}, 477 (1995).

\bibitem{bi:Kofman}
L. Kofman, E. Bertschinger, J.M. Gelb, A. Nusser and A. Dekel,
Ap.J {\bf 420}, 44 (1994).

\bibitem{bi:Vachaspati}
T. Vachaspati, Phys.Lett.B {\bf 282}, 305 (1992).

\bibitem{bi:Moessner}
R. Moessner, MNRAS {\bf 277}, 927 (1995). 

\bibitem{bi:Martins}
C.J.A.P. Martins, Phys.Rev. D {\bf 55}, 5208 (1997).

\bibitem{bi:vdBruck}
C. van de Bruck, astro-ph/9705208, submitted to Phys.Rev. D.

\bibitem{bi:MartinsShellard}
C.J.A.P. Martins and E.P.S. Shellard, Phys.Rev.D {\bf 54}, 2535 (1996).

\bibitem{bi:albrecht}
A. Albrecht, R.A. Battye, J. Robinson, astro--ph/9707129 (1997)

\bibitem{bi:avelino}
P. Avelino, R.R. Caldwell and C.J.A.P. Martins, to appear in Phys.Rev.D, 
astro-ph/9708057 (1997).

\bibitem{bi:vachiperi}
T. Vachaspati and L. Perivolaropolous, ApJ {\bf 123}, 456 (1994).

\bibitem{bi:hara}
T. Hara. P. M\"ah\"onen and S. Miyoshi ApJ {\bf 415}, 445 (1993)

\bibitem{bi:zeldovich}
Y. B. Zeldovich, A \& A {\bf 5}, 84 (1970)

\bibitem{bi:peri}
L. Perivolaropoulos, Phys. Lett. {\bf B298}, 305 (1993)

\bibitem{bi:Peebles}
P.J.E. Peebles, The large scale structure of the universe, 
Princeton Univ. Press (1980).

\bibitem{bi:Dekel}
A. Dekel, ARAA {\bf 32}, 371 (1994)

\bibitem{bi:Babul}
A. Babul, D.H. Weinberg, A. Dekel and J.P. Ostriker, ApJ {\bf 427}, 1 (1994)

\end{thebibliography}
\end{document}